\documentstyle[prb,aps,epsf,floats]{revtex}
\newcommand{\beq}{\begin{equation}}
\newcommand{\eeq}{\end{equation}}
\newcommand{\beqr}{\begin{eqnarray}}
\newcommand{\eeqr}{\end{eqnarray}}

\def\bn{{\mathbf n}}

\def\bS{{\mathbf S}}

\def\bL{{\mathbf L}}

\newcommand{\sigmab}{\mbox{\boldmath $\sigma $}}

\newcommand{\p}{{\partial}}

\def\eqa{\begin{eqnarray}}
\def\eea{\end{eqnarray}}

\begin{document}
\draft \flushbottom \twocolumn[
\hsize\textwidth\columnwidth\hsize\csname
@twocolumnfalse\endcsname
\title{ Deconfinement in $d=1$: a closer look.}
\author{R. Shankar$^1$ Ganpathy Murthy$^2$, }
\address{$^1$ Department of Physics, Yale
University, New Haven CT 06520\\ $^2$Department of Physics and
Astronomy, University of Kentucky, Lexington KY 40506-0055 }
\date{\today}
\maketitle
\begin{abstract}
The notion of deconfinement in two $d=1$ models, the  Schwinger
model and the Heisenberg chain,  is re-examined. Both have
half-asymptotic excitations (where particles and antiparticles must
alternate)  and also truly asymptotic particles which are half as
many in number. The two kinds of particles are  related by a
complicated transformation. The main purpose of this note is to
highlight the relationship between asymptotic and half-asymptotic
particles. The relevance of our findings to higher dimensions is
briefly discussed.
\end{abstract}
\vskip 1cm \pacs{XXXX}]
\section{Introduction}
\label{intro}

We take a closer look at deconfinement  in   $d=1$ and point out
that
 the situation is more complicated than is generally appreciated.  We
 consider   two examples, the massive Schwinger model and the Heisenberg spin chain,
  wherein  the degrees of freedom one
 often considered as deconfined
 are  really {\em  half-asymptotic}.  That is, particles and antiparticles can move arbitrarily far from each other,
 provided they   alternate along the line.
 In both cases however, there exist truly  asymptotic, unrestricted
 particles,
 but these are fewer in number and related
  to the half-asymptotic  ones by complicated
 transformations.

  Many results presented
  here are scattered in the literature, and we claim credit only
 for filling in some gaps and highlighting  the distinction between asymptotic and half-asymptotic particles.

\section{Example 1: The massive Schwinger model  }
Schwinger posed and solved the eponymous model of the
electrodynamics of massless Dirac fermions of charge $e$ and showed
that the final spectrum consisted only of a single boson of  mass
${e\over \sqrt{\pi}}$. This result, readily established using
bosonization, is understood as the confinement of charged particles
by the linearly rising Coulomb potential (which is what
electrodynamics reduces to in one space dimension).  The case with
non-zero fermion mass $m$ was studied by numerous authors, most
notably Coleman  and coworkers\cite{coleman}.  They pointed out that
electrodynamics in $d=1$ admits an extra parameter $\theta$ that
corresponds to a background electric field $E_{\theta}=e{\theta
\over 2\pi}$ produced by charges $\pm e{\theta \over 2\pi}$ at
$x=\mp \infty$.  In the absence of matter fields, the field  lines
due to these boundary charges would run undiminished in density from
$-\infty$ to $+\infty$. In the massless case any such field would be
screened immediately and at no cost by the fermions via what is
essentially Debye screening. Thus $\theta$ would be effectively
zero. (In the bosonized version, $\theta$ can be shifted away. ) In
the massive case, a suitable integer part of the background charges
would be screened by the fermions, leaving a background field
obeying $-{e\over 2}< E_{\theta} \le {e\over 2} $, that is with
$\theta ={2\pi E_{\theta}\over e}$ limited to $-\pi< \theta \le
\pi$. Note that even though massive fermions cost a finite amount of
energy, they are worth producing if bulk energy density can be
reduced.

 Let us consider the phase diagram of this model as a
function of $\theta$, $m$,  and $e$. We study the weak coupling
limit in the fermionic version and then pass on to the bosonized
version for strong coupling. We will begin by simply reviewing
Coleman's work. Our contributions, pertaining  to the interpretation
and clarification of the phase transitions (and its relation to our
main theme) will then follow.

The hamiltonian is \begin{eqnarray} H&=& \int \psi^{\dag}(\alpha p
+\beta m)\psi  dx  \nonumber \\ & &  - {e^2 \over 4} \int \int \rho
(x) |x-y|\rho (y) dx dy - {e^2 \theta \over 2 \pi}\int x \rho (x) dx
\label{diracham}
\end{eqnarray} where $e \rho$ is the charge density and $\alpha $
and $\beta $ are Dirac (Pauli) matrices. The first term is the
fermion kinetic-energy, the second is the electrostatic energy of
dynamic charges, and the third is the coupling of the matter fields
to the background field. (The energy to set up the background field
is not shown.)

Gauss' law relates the electric field $E$ {\em due the dynamic
charges} as per \beq
 {dE\over dx}=e \rho .\eeq
Note that this $E$ is the field due to the fermions only and does
not include the background field. However, the divergence of the
total field $E_T$ is the same as that of $E$.

 Using Gauss' law  can rewrite the last term in Eqn. (\ref{diracham}) as
follows: \beq -{e \theta \over 2 \pi}\int x {dE (x)\over dx}
dx=E_{\theta}\int E(x)dx \label{topo} \eeq where $E_{\theta}=
e\theta /2\pi.$ Indeed,  the total electrical energy density
${1\over 2}E_{T}^2={1\over 2}(E+E_{\theta})^2$, generates  a term
 $E_{\theta}^2/2$ that is suppressed, a term $E^2/2$ which is the second term in
Eqn. (\ref{diracham}) and cross term which gives the the last term
 $E_{\theta}E$. It will be useful to remember for
later use that the $\theta$ term is the integral of $E_{\theta}E$.

Let us start with  empty space permeated by the background field
$E_{\theta}$ and $\theta$ chosen to be positive. (Negative values
lead to the same physics upon making suitable sign changes.) Suppose
we introduce a fermion-antifermion pair $F$-$\bar{F}$ with the
antifermion to the left.
 We see that $E_T$ starts at the value ${e\theta \over 2\pi}$ at $x=-\infty$
  and drops down to $e({\theta \over 2\pi}-1)$ between the  charges and goes back
  to ${e\theta \over 2\pi}$ to the right of
 $F$. The reader can verify easily that in the region between
 charges the there is an extra energy per unit length $\delta U=
 1-{\theta \over \pi}>0$. Thus charges are linearly confined in a generic case.
 (Had the $F$ occurred to the left,  the field between $F$ and $\bar{F}$ and  confining potential would have
 been
 even larger.)

 However at $\theta =\pi$, the charges can be
 separated with no extra cost (ignoring residual short range forces) {\em provided $\bar{F}$ lies to the
 left of $F$.} (A similar thing happens at  $\theta =-\pi$, where the  external field
 is reversed, provided we reverse the  ordering of $F$ and $\bar{F}$.) We can introduce any
 number of such pairs at finite cost provided $F$ and $\bar{F}$ alternate.

 The particles at $\theta =\pi$ were described as {\em half-asymptotic} by
 Coleman. These are the naive "spinons" of this problem. They do
 not constitute genuine deconfined  degrees of freedom because of the restriction that fermions and antifermions alternate.
  For example,
 if we are  sitting next to a fermion, millions of miles away from any other matter, we know
 that on either side of it must lie antifermions.
   What we need are
 objects that can be created with no such restriction. We will find them,
 but they will be related to  $F$ and $\bar{F}$ in a
 very complicated way. There will also be half as many of them.

 To proceed we must bosonize. For our purposes, bosonization is a set of
 rules \cite{colemanboso,kathmandu,fradkin,subir,thiery} given by:
 \begin{eqnarray}
\psi^{\dag} \alpha p \psi &=& {\pi^2+ (\nabla \phi )^2 \over 2}\\
\psi^{\dag}\beta \psi &=& -c\Lambda \cos (\sqrt{4\pi}\phi)\ \ \
\ \ \ \\
 \psi^{\dag}\psi &=& \rho
={1\over \sqrt{\pi}}{\p \phi \over
\partial x}
\end{eqnarray}
where $\Lambda $ is the cut-off, and c is a constant.  Comparing the
last equation to Gauss's law we see that $\phi$ is essentially the
electric field due to the dynamic charges: \beq E={e\over
\sqrt{\pi}}\phi \eeq and that the net fermion charge is proportional
to $\phi (\infty )-\phi (-\infty )$. We will limit ourselves to the
neutral sector so that we may choose $\phi (\infty )=\phi (-\infty
)=0$.

The bosonized hamiltonian becomes, upon doing some integrations by
parts (recalling  that $\phi$ vanishes at infinity),  defining $\phi
+ {\theta \over 2 \sqrt{\pi}}$ as the new $\phi$, and using
$\nabla^2 |x-y|=2\delta (x-y)$, \beq H=\int dx \left[ {\pi^2+
(\nabla \phi )^2 \over 2}+{e^2\over 2\pi}\phi^2 - mc\Lambda \cos
(\sqrt{4\pi} \phi - \theta)\right] . \eeq Note that due to the
shift,  $\phi$ now is proportional to the {\em total electric field
$E_T$} due to external and dynamic charges and that in the massless
case $m=0$, $\theta$ drops out.

 Let us examine the scaled potential energy \beq v(\phi )={V(\phi
)\over cm\Lambda}= \lambda \phi^2 + \cos (\sqrt{4\pi} \phi +
\gamma)\eeq where \beq \gamma = \pi -\theta \eeq and \beq \lambda =
{e^2 \over 2\pi m \Lambda c}.\eeq

First consider weak coupling $\lambda <<1$. Start with $\gamma =0$.
The cosine has an infinite number of minima. Focus on the ones
nearest to the origin, near $\phi =\pm {\sqrt{\pi}\over 2}$, since
the other minima will be lifted to  higher  energies when we turn on
the $\lambda \phi^2$ term. These two values of $\phi$ correspond to
total electric fields $\pm e/2$. How do we get two values for $E_T$?
The value $e/2$ corresponds to the background field at $\theta =\pi$
while the one at $-e/2$ corresponds to the case when an $\bar{F}$
and $F$ have been produced and sent to $\mp \infty$ (at zero energy
cost {\em per unit volume}) to produce a degenerate configuration.
For a small $\gamma
>0$, there is just one minimum near $\phi = \sqrt{\pi}/2$. As the
potential $v(\phi )$ has a unique minimum, there are no solitons and
hence no fermions. This is the effect of linear confinement at
$\theta \ne \pi$. As $\gamma$ passes through zero, to negative
values, $\langle \phi \rangle $ jumps to  roughly  $ -
\sqrt{\pi}/2$. The transition at  $\gamma =0$ is thus first order.
(When $\gamma$ becomes negative,  $\theta $ exceeds $\pi$, at which
point an $F-\bar{F}$  pair is sent to spatial infinity to bring
$\theta$  to the  allowed interval $|\theta |\le \pi $. )

Let us now focus on the critical line $\gamma =0$ (that is $\theta
=\pi$) and slowly raise $\lambda$. At small $\lambda$, we have two
degenerate minima in $\phi$  near $\pm \sqrt{\pi}/2$.  Solitons
connecting these two vacua can be identified with the old $F$ and
$\bar{F}$. However since we have only two minima, an increase in
$\langle \phi \rangle $ must be followed by a decrease in $\langle
\phi \rangle $ and vice versa, i.e., particles must alternate with
antiparticles. These solitons and antisolitons are the
half-asymptotic particles.

At some very large  value of $\lambda$ (or $e^2$), the curvature of
the bare potential $v(\phi)$ at the origin is positive and there is
just one ground state at $\phi =0$. There are only neutral bosons
beyond this value of $\lambda$, and no  half-asymptotic fermions.
Due to quantum fluctuations the phase transition will occur at a
smaller value, $\lambda_C$. The phase diagram is then as shown in
Figure 1. This concludes the review of past work.

\begin{figure}
\narrowtext \epsfxsize=2.4in\epsfysize=2.6in \hskip
0.3in\epsfbox{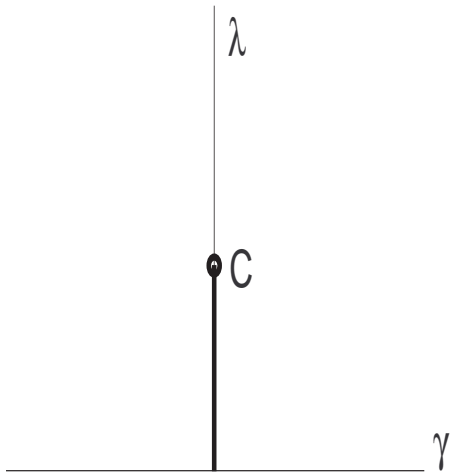} \vskip 0.15in \caption{Phase diagram in the
$\lambda ={e^2 \over 2\pi m \Lambda c}$  and $\gamma=\pi - \theta $
plane. The dark line denotes a first order transition terminating at
$C$, a second order critical point. In the Ising model mapping,
$\lambda -\lambda_C$ plays the role of $t$, the temperature and
$\gamma = \pi - \theta $ plays the role of the magnetic field $h$.}
\label{scatt-amplitude1}
\end{figure}

We now ask about the nature of $C$,  the end point of the line of
first order transitions. As the only symmetry in the problem is
$\phi \to -\phi$,   the transition is generically in the Ising
class. Clearly $\lambda -\lambda_C$ plays the role of temperature
$t$ and $\gamma = \pi -\theta $ plays the role of the symmetry
breaking magnetic field $h$. Let us discuss this in more detail. For
small $\lambda$, i.e., deep in the ordered side, the low energy
configurations are a series of kinks as the ground state jumps from
one value of $\langle \phi \rangle $ to its negative. We can then
describe the kinks by domain wall fermion operators $c^{\dag},c$.
(They are fermionic because there can only be zero or one wall at a
site.) These walls can move and also annihilate in pairs. It is well
known that this wall dynamics describes the Ising model in operator
form (using the transfer matrix). Equivalently, we can associate an
Ising spin $\sigma_3$ with the two vacua (call them "up" and "down"
vacua) and describe the low energy physics with the Ising
hamiltonian \beq H_{I }= - \sum_n \sigma_3(n)\sigma_3(n+1)- (1+t)
\sum_n \sigma_1(n) \eeq where the first term assigns a cost to
domain walls and the second term allows spin flip, i.e., vacuum
fluctuations and $t=0$ is the critical point. Let us define  two
Majorana fermions $\psi_1$ and $\psi_2$ (related to the hermitian
operators $c+ c^{\dag}$ and $i(c-c^{\dag})$) by the strings \beq
\psi_{1,\ 2}(n)={1 \over
\sqrt{2}}\bigg(\prod_{-\infty}^{n-1}\sigma_1(k)\bigg)\sigma_{2,3}(n)\
)\label{majorana} \eeq and which obey the anticommutation rules \beq
\left[ \psi_{\alpha}(n),\psi_\beta(m)\right]_+=\delta_{\alpha
\beta}\delta_{mn }.\eeq \ In terms of these \beq H_{I}=-2i\left[
\sum_n \psi_1(n)\psi_2 (n+1) +(1+t)\sum_n \psi_1(n)\psi_2(n)\right].
\eeq

It is these Majorana fermions that are truly liberated at the first
order transition $\gamma =0$.  They can be created at will. Unlike
the Schwinger fermions or antifermions that {\em shift} $\phi$ in
one direction or another in a half-space, the Majorana fermions {\em
reflect} $\phi$ to minus the value in a half-space. While the
half-asymptotic fermion creation operator of the Schwinger model
only act on the "down" vacuum sending it to "up" in half of space
(and the antifermion does the opposite), the Majorana fermion can
act on either vacuum and flip it to the other. We may schematically
represent these two kinds of fermions by the strings

\begin{eqnarray}
\Psi_{Schwinger}^{\dagger} &\simeq &\prod_{-\infty}^{n-1} \sigma_+(k)\\
\psi_{Ising-Majorana} & \simeq &\prod_{-\infty}^{n-1} \sigma_1(k)
\end{eqnarray} It is clear that there is no simple relation between
the two.  To summarize:
\begin{itemize}
\item The end point of the first-order transitions at $\gamma =0 \
\mbox{or} \ \theta =\pi$ is an Ising transition, the point C in
Figure 1. \item The charged fermions of the Schwinger model are
deconfined at the first-order transition at $\theta =\pi$ but are
half-asymptotic, while the Majorana fermions are truly deconfined
objects and correctly describe the second order end point. The
Majorana fermions, being domain walls, are also confined away from
the line $\gamma =0$ since the spins in one side or other of a
domain wall  are pointing opposite  to the applied field.  \item The
number of truly deconfined objects is half that of the
half-asymptotic particles.
\item There is no simple relation between the two kinds of particles.
\end{itemize}

Note that the fact that Majorana fermions, related to domain walls
of the ordered side,  also describe the disordered side (where
$\langle \phi \rangle =0$). This is due to the self-duality of the
Ising model: if we formed Majorana  kinks out of the disorder
variables, \beq \mu_1 (n)  = \sigma_3(n)\sigma_3(n+1)\ \ \ \
\mu_3(n)= \prod_{-\infty}^{n} \sigma_1 (k)\eeq
 we would get essentially the same operators as in
 Eqn.(\ref{majorana}) except for the exchange between
 $\psi_1$ and $\psi_2$. Of course, these fermions would cease to
 exist deep in the disordered phase when $\lambda$ exceeds the value
 beyond which the bare potential $v(\phi)$ has only a single minimum
 at $\phi=0$.

\section{Example II: The spin chain and spinons}

One place where spinons arise is in the study of the following
hamiltonian of spin-${1\over 2}$ degrees of freedom studied in depth
by Haldane \cite{haldane}:
\begin{eqnarray} H&=&J_1\sum_n \bS (n)\cdot \bS (n+1) +J_2\sum_n  \bS
(n)\cdot \bS (n+2)\nonumber\\ &+&\gamma \sum_n (-1)^n \bS (n)\cdot
\bS (n+1).\label{spinham}\end{eqnarray} Let us rescale $J_1$, the
coefficient of the Heisenberg interaction, to unity. With only this
term, the model is exactly solvable and known to have power law
decay of all correlations. We are interested in the
case\cite{haldane} wherein $J_2>J_{2c}$, when the system exhibits
spontaneous dimerization, that is \beq  \left< (-1)^n \bS (n)\cdot
\bS (n+1)  \right> =\pm \Omega_{SP} ,\eeq where "SP" denotes the
degenerate Spin-Peierls ground states. Let us call them "up" and
"down" vacua respectively. In such a state the bond energy between
neighboring spins alternates in strength as we move across the
lattice. In the last term, the coefficient $\gamma$ is an external
field that couples to this order parameter and chooses one of the
two degenerate states. {\em It will be turned off in what follows
unless otherwise stated. } It is known that the case $J_2=.5 J_1$,
the Majumdar-Ghosh hamiltonian\cite{mg}, lies in this dimerized
phase. At this point $H$ is a sum of operators that project any
three spins into a triplet state. It follows that when any two of
the three form a singlet it is annihilated by this projector. The
two ground states then are of the  form \beq |{down}\rangle
=...B_{12}B_{34}B_{56}...\eeq \beq |{up}\rangle=
..B_{23}B_{45}B_{67}..\eeq where $B_{ij}$ is a singlet formed out of
the spins at sites $i$ and $j$. In the up (down) state the strong
bond has an even (odd) numbered site at its left end.

A spinon at $n=0$ is imagined as follows. From $-\infty$ to $n=-1$
we have the up vacuum, at $n=0$ there is a free (unpaired) spin,
followed by the down vacuum whose strong bond begins at $n=1$.
Further down the chain we can insert an anti-spinon, but it has to
be at an odd site so we can switch back to the up vacuum. It appears
that there are four states open to a spinon: it can have its spin up
or down and it can occur as we go from the up to down vacuum or the
other way around. Note however that an anti-spinon that switches
from up vacuum to down must be followed by a spinon  that switches
from down to up. This is alternation is reminiscent of the
$F-\bar{F}$ pairs in the Schwinger model. We shall see once again
that compared to these degrees of freedom that are half-asymptotic,
the number of really independent deconfined degrees of freedom are
half as many. Note that the third term in Eqn. (\ref{spinham}),
(proportional to $\gamma$) if turned on, will lead to a confinement
of spinons just like a magnetic field confines domain walls in the
Ising model. The half-asymptotic spinons are thus deconfined at a
first order transition when $\gamma$ vanishes.

We shall exhibit the half-asymptotic nature of these spinons in
another language which parallels the Schwinger model  more closely.
First we begin with Haldane's mapping \cite{haldanespinchain} of the
spin chain to the nonlinear sigma model with the following
Lagrangian density: The Lagrangian density in terms of the unit
three-vector representing N\'eel order $\bn$ is \beq {\cal L}=
{1\over 2g^2}(\p_{\mu} \bn \cdot \p^{\mu}\bn)+i\lambda (|\bn
|^2-1)+{i\theta \over 4\pi} \bn \cdot {\partial \bn \over
\partial x }\times  {\partial \bn \over
\partial \tau }.\eeq
 The vector
$\bn$ is normalized to unity by the Lagrange multiplier field
$\lambda$ and $g$ is the coupling constant related to the
microscopic spin hamiltonian in some manner that does not concern
us. The parameter $\theta$ equals $\pi$ when $\gamma$ =0.

 This model in
turn can be mapped into the $CP^{1}$ model\cite{dadda} with the
Euclidean Lagrangian density
\begin{eqnarray} {\cal L} &=& \int d^2x \left[{2\over g^2}|D_{\mu}z|^2+i \lambda (\bar{z}z-1)+{i\theta \over
2\pi}\varepsilon_{\mu \nu} \p_{\mu}a_{\nu} \right]\\
D_{\mu}&=& \p_{\mu}-ia_{\mu}\ \ \ \ \ a_{\mu}=-i\bar{z}\p_{\mu}z
\end{eqnarray}
where $z$ is a  2-component spinor,  related to  $\bn$ via \beq \bn
= \bar{z}\sigmab z. \eeq Note that the gauge field $a_{\mu}$ does
not have any dynamics at this stage.

The equality of the two Lagrangians can be shown by using Fierz
identities such as \beq \sigmab_{ab}\cdot \sigmab_{cd}=
-\delta_{ab}\delta_{cd}+2\delta_{ad}\delta_{bc}.\eeq

 The relation $\bn =\bar{z}\sigmab z$
implies that $z$ may be rotated in phase from point to point, which
is why we get a gauge theory. The reason such gauge theories are
attractive is that spinor excitations appear naturally in them. On
the other hand, spinor excitations are difficult, if not impossible,
to construct if one works directly with the vector $\bn$.

If $z$ had $N$ components instead of 2, with $N\to \infty$, we would
proceed as follows\cite{witten}:
\begin{itemize}
\item Integrate over the $z$ quanta of which there are $N$
species. \item Find the saddle point of the effective action with
an $N$ in front of the trace-log coming from the integration.
\item Note that $\lambda$ has a mean value (which gives a mass $M$
to the $z$) and negligible fluctuations. \item Note that the field
$a_{\mu}$ acquires dynamics, i.e., has an action proportional to
$F_{\mu \nu}^{2}$ which mediates a linear potential between the
z-quanta.
\item Observe that the $\theta$ or topological term corresponds to a background field with
charges ${\theta \over 2\pi}$ times the $z$-quanta charge. One way
to see this is to write the term as a line integral around the
boundary of space-time using Stokes' theorem. At any given time
slice we intercept  two Wilson lines of charges $\pm \theta /2\pi$
at spatial infinity. From our discussion of this term in the
Schwinger model we see that $\varepsilon_{\mu \nu} \p_{\mu}a_{\nu}$
is the electric field of the matter charges.
\end{itemize}

While the large $N$ picture is incorrect for $N=1$ if $J_2$ is
small, it correctly describes  the dimerized phase, $J_2>J_{2c}$
because here the $z$-quanta are massive, as assumed in the
derivation of the effective action for $a_{\mu}$.  The effective
theory at low energies is thus given by the Lagrangian density \beq
{\cal L}=|(\p_{\mu} -iea_{\mu})z|^2-M^2 \bar{z}z-{1\over 4}F_{\mu
\nu }^{2}+{ie\theta \over 2\pi}\varepsilon_{\mu \nu}
\p_{\mu}a_{\mu}.\eeq
The $a_{\mu}$  field has been rescaled to get
to this form \cite{witten}. Note that in the effective theory,
$a_{\mu}$ is an independent field, not slaved to $z$,  and $e$ is a
 parameter arising during $z$- integration.

We see that this is just the Schwinger model  with the trivial
modification that the particles and antiparticle carry spin and are
bosons and the global symmetry group is SU(2). The $\theta$ term can
be viewed as the product of the external field $E_{\theta}$ due to
boundary charges $\pm {e\theta \over 2\pi}$ and the electric filed
due to dynamic charges, $E= \varepsilon_{\mu \nu} \p_{\mu}a_{\mu}$.
 At $\theta =\pi$ (or $\gamma =0$) the total field $E_T$-field reverses sign
whenever one crosses a particle or antiparticle  since the particle
charge is double the external charges at $\pm \infty$.

There is thus very strong circumstantial evidence that the  $z$ and
$\bar{z}$ quanta are  the half-asymptotic spinons previously
caricatured  as the dangling spins that separated the two dimerized
states with opposite values of $\Omega_{SP}$. To firm up  the
connection we need to show that the   electric field
$\varepsilon_{\mu \nu} \p_{\mu}a_{\mu}$ which goes up  every time we
cross a $z$ or goes down when we cross a  $\bar{z}$ is none other
than the order parameter density $(-1)^k \bS (k)\cdot \bS ( k+1)$
that does the same  when we cross a spinon or antispinon.

We offer two proofs. First, we can argue that if we change the
Lagrangian  by adding a term proportional to the integral of
$\varepsilon_{\mu \nu} \p_{\mu}a_{\mu}$ this changes $\theta$ away
from $  \pi$ and causes linear confinement, which is exactly  what
happens if we add a term proportional to the Spin-Peierls order
parameter, i.e., turn on $\gamma$ in Eqn. (\ref{spinham}).

A more microscopic answer can be given to readers  more familiar
with the sigma model description of the spin chain in terms of the
unit three- vector $\bn = \bar{z}\sigmab z$ and the associated
angular momentum $\bL=\bn \times {\p \bn \over \p t}$. Starting with
Haldane's decomposition \cite{haldanespinchain} of $\bS$,  in terms
of $a$, the lattice spacing, $\bL$ and $\bn$,
\begin{eqnarray} \Omega_{SP}&=&  (-1)^k \bS (k)\cdot \bS
(k+1)\nonumber \\ &= &  (-1)^k (a\bL (k) +(-1)^k\bn (k))\cdot (k \to
k+1 )\nonumber  \\ &= &   \! \! a \bL
(k)\cdot (\bn (k) -\bn (k+1)) +\mbox{ oscillating terms}\nonumber  \\
&= & -a^2 \bL \cdot \nabla \bn \to a^2 \bn \cdot {\p   \bn \over \p
x} \times {\p \bn \over \p t}\nonumber ,\end{eqnarray} which is
proportional to the topological density in the $\bn$ language, which
in turn is given in the $CP^{1}$ language by $\varepsilon_{\mu
\nu}\p_{\mu}a_{\nu}$. Thus we have related $\Omega_{SP} $ to the
electric field of the $CP^{1}$ model.

If the $z$-quanta are half-asymptotic what are the really deconfined
particles and how many of them are there? Once again bosonization
provides the answer, now applied to the   Jordan Wigner fermions
formed from spin operators. The details will not be furnished here
since they are copiously described in the published literature, see
for example \cite{haldane,subir,thiery}.

The bosonized Lagrangian density is \beq {\cal L}=\alpha
((\p_{\tau}\phi )^2+ (\nabla \phi )^2)-v \cos (4\phi ) \eeq where
$\alpha$ and $v$ are functions of $J_2$ (and $J_1$ which can be set
to unity by overall rescaling). The normalization of the field $\phi
$ is as in Sachdev's book \cite{subir} and but not the same as the
one used in the Schwinger model. Space-time anisotropy has also been
removed by a suitable rescaling of coordinates. The only important
points are the following. \begin{itemize} \item  The coupling  $v$
becomes negative for $J_1>J_{2c}$., when we enter the Spin-Peierls
state. This is the regime of interest to us. \item The boson field
is related to the underlying spin variables as follows:
\begin{eqnarray}
\sin (2\phi )&=&  (-1)^n \bS (n)\cdot \bS (n+1)\equiv \Omega_{SP}\\
\cos (2\phi ) &=&(-1)^n S_{z} (n)\equiv \Omega_{N} \\
\sum_k S_z (k) &= & {\phi (\infty )-\phi (-\infty )\over 2\pi}
\end{eqnarray}
where the labels $N$ and $SP$ on the order parameters  stand for
Spin-Peierls and N\'{e}el respectively.
\item The underlying spin variables are invariant under $\phi \to \phi +\pi$. Thus
for example, even though there may seem to be an infinite number of
vacua in $\phi$ they describe a finite number of physically distinct
situations.
\end{itemize}

Let us examine the vacuum and soliton structure of the potential
energy

\beq v(\phi )=|v|\cos (4\phi )\eeq whose minima are at \beq
\phi^*=(2m+1){\pi \over 4}. \eeq At these minima $\Omega_{N}$
vanishes while \beq \Omega_{SP} = \pm 1 \ \  \mbox{according as $m$
is even/odd} .\eeq Let us again refer to these vacua where
$\Omega_{SP}=\pm1$ as up and down vacua. Note that there are just
two kinds of vacua here, depending on whether  $m$ is even or odd.
In other words even though $m$ has an infinite range of values,
physics  depends on $m$ {\em modulo 2}.

Suppose we start out with $m=0$ (up vacuum) at $x=-\infty$. We can
go to the down vacuum either by going to $m=-1$ or $m=+1$. These two
values of $m$ differ by $2$, the corresponding $\phi$'s differ by
$\pi$, which makes  these vacua physically equal as far as  the
spins go. However these solitons  carry spins
 $S_z=\mp {1\over 2}$ depending on which way we move in $\phi$. These are the two
spin states of the spinon  that takes an up vacuum to a down vacuum.
However no distinction is made between this and a spinon that
connects a down vacuum to an  up vacuum, one that starts out at say
$m=1$ and ends up at $m=0$ or $m=2$ corresponding to $S_z=\mp
{1\over 2}$. That is, these domain walls are once again neutral.
There is just one kind of domain wall here, it carries spin and
separates the two vacua, and the same domain wall implements an up
to down or down to up transition.

On considering the partition function of the lattice $CP^{1}$ model
invented by Seiberg \cite{seiberg}, at $\theta =\pi$, Affleck
\cite{iancp} has noticed that since particles and antiparticles have
to alternate (if we want zero coulomb energy), the charge label is
not needed and the partition function, which is a sum over loops
reduces to that of a system of neutral particles, i.e., with no
orientation. He comments that (for general $n$ in the $CP^{n-1}$
model) the $SU(n)$ model seems to have reduced to an $O(n)$ model.
We emphasize that despite this reduction upon the restriction of the
full model to the zero coulomb energy sector, the symmetry of the
$CP^{1}$ model is firmly anchored at SU(2), with the spinons with
$S_z=\pm {1\over 2}$ forming an isospinor.

\section{Conclusions and outlook}
We re-examined deconfinement in  $d=1$, which is after all the
source of many of the most fruitful ideas in the physics of strongly
correlated electrons. We found that the situation was more complex
than is generally appreciated. In the two cases we studied, the
massive Schwinger model and dimerized spin chain, the particles
often viewed as deconfined were only half-asymptotic (had to be
created with particles and antiparticle alternating). However
 in both cases there were also truly deconfined
asymptotic particles, but these were half-as many in number and
related in a complicated way to the more obvious but half-asymptotic
ones. In both the Schwinger model and  the spontaneously dimerized
spin-1/2 chains there were two vacua, called up and down,  and the
half-asymptotic particles and antiparticles were associated with the
passage from down to up or up to down vacua respectively. The truly
asymptotic ones were associated with domain walls that simply lay
between any two vacua, that is say, they seemed "neutral". In the
Schwinger model the original fermions and antifermions were
half-asymptotic on critical line $\theta=\pi$ while the Majorana
fermion was truly asymptotic and described the Ising transition. In
the magnetized problem,  there were two kinds of domain walls going
from down to up and up to down vacua, each with two values of spin.
In the $CP^1$ version we had particles and antiparticles, namely $z$
and $\bar{z}$ quanta, each with two values of spin.  Both were
half-asymptotic. The truly asymptotic particles however only carried
spin and could connect any one vacuum to the other.

In these examples the doubling occurred only in the gauge
description of the problem. Why bother to set up a straw man and
shoot him? Why not just limit ourselves to to the non-gauge
descriptions that did not have this redundancy? The answer is that
the very notion of deconfinement exists only in the gauge version
and gauge models are the focus of a lot of attention, and rightly
so,  since they allow one to introduce fractionalized objects into
the theory.

It is interesting to ask how  the Ising model with its domain walls
could be mapped into  a gauge theory.  We first attach an extra
label to the domain walls calling the ones interpolating from down
spin to up spin vacua as particles those that do the reverse as
antiparticles. However we must now demand that particles and
antiparticles alternate. One way to enforce this is to couple them
to a gauge potential that produces linear confinement, introduce a
background field corresponding to $\theta =\pi$, and let them
deconfine as half-asymptotic particles forced to alternate in the
low energy sector, i.e, end up with the massive Schwinger model.

One might think that the factor of two redundancy in the gauge
version is due to the usual redundancy of gauge descriptions, i.e.,
because  the Schwinger fermions and the z quanta are not gauge
invariant. To see that this is not so one need only consider QED in
the coulomb phase, say in three dimensions: The electron and
positron operators are   not gauge invariant, but these particles
are very much part of the spectrum. The reduction in true degrees of
freedom arises because  we want to limit ourselves to the low energy
sector, with no coulomb energy. In the Schwinger model for example,
we can have fermions and antifermions in any order if we are willing
to go to states of high electrostatic energy. A better analogy is to
consider the restriction to the lowest landau level in the FQHE
problem. This  restriction changes a lot of commutation relations,
$x$ and $y$ become conjugate, the $x-y$ plane becomes the phase
space and the problem becomes one-dimensional.

To better understand the impact on physics in $d>1$, we need to
recall the history. This has to be necessarily brief given the
rather long time separating today from Anderson's  proposal
\cite{pwa} that spinons that appear in $d=1$ could also appear in
$d=2$. For this to happen, the spin system must be in a featureless
liquid state. Rokhsar and Kivelson \cite{rk} introduced a model of
hard core quantum dimers (representing spin singlet nearest neighbor
bonds) which they solved exactly at a particular point (known now as
the RK point). At the RK point spinons are indeed deconfined. Read
and Sachdev \cite{readandsachdev}, using large $N$ generalizations
of SU(2) spins, found that the systems like to either have N\'eel
order or Valence Bond Solid (VBS) order. They also pointed out that
the Rokhsar-Kivelson model had order on both sides of the RK point,
implying that it was an isolated point of deconfinement. This fact
was made explicit by Moessner, Sondhi, and Fradkin \cite{msf}, who
appealed to the quantum Lifshitz theory introduced by
Henley\cite{quantum-lifshitz}. It turns out that the RK point is a
result of fine tuning (in terms of underlying spins) and the
deconfinement at criticality it exhibits is not generic\cite{ash}.
In the meantime, an entire deconfined phase  has been uncovered
\cite{deconphases} by Moessner and Sondhi in the triangular lattice
hard-core dimer model. Fractionalized deconfined phases have also
been proposed in other contexts as well \cite{qd-models-gauge2}.
Most recently, the idea that deconfinement at criticality could be
generic in a class of spin models that also exhibited non-Landau
phase transitions (where two different order parameters vanished
from either side) has been vigorously pursued by Senthil {\em et
al}\cite{senthil}.

 Can our study here shed any light on these works in $d>1$.
While it is true that in $d>1$, as in $d=1$, spinons are confined by
gauge fields, it does not appear that there is any double counting
either in the caricatures in terms of spins or gauge theory
descriptions.
 The mechanism for deconfinement
in $d=1$ (an external field getting reversed by the field due to the
particles without bulk energy cost ) does not seem applicable in
$d=2$. In other words, it seems as if particles will be  either
confined or librated in $d=2$ with no room for half-asymptotic
particles, the very  notion being limited to $d=1$ where particle
ordering makes sense. Furthermore, the deconfining transitions we
considered were generically first order, while the focus in $d>1$
has been in continuous second-order transitions. The only common
feature may be that the relation between deconfined particles and
their confined versions on either side of the transition is
hopelessly complicated.

We are grateful for many illuminating conversations with I. Affleck,
Vadim Oganesyan, Subir Sachdev, T.Senthil, Shivaji Sondhi,   Ashwin
Vishwanath and Xiao-Gan Wen.
 We acknowledge the National Science Foundation  for grants DMR  0354517  (RS) and  DMR 0371611    (GM) and the Aspen
 Center for Physics where this work was completed.

 \end{document}